\documentclass[12pt,preprint]{aastex}
\usepackage{epsfig}
\usepackage{rotating}
\usepackage{lscape}

\shorttitle{Water megamaser emission in submm galaxies}
\shortauthors{Wagg and Momjian}
\slugcomment{AJ accepted}

\def\ga{\mathrel{\raise0.35ex\hbox{$\scriptstyle >$}\kern-0.6em
\lower0.40ex\hbox{{$\scriptstyle \sim$}}}}
\def\la{\mathrel{\raise0.35ex\hbox{$\scriptstyle <$}\kern-0.6em
\lower0.40ex\hbox{{$\scriptstyle \sim$}}}}

\def\cothree{CO~{\it J}=3-2 }

\def\hij{high-{\it J}~}

\def\smmjof{SMM~J14011+0252~}
\def\smmjos{SMM~J16359+6612~}

\begin{document}

\title{Constraints on the presence of water megamaser emission
  in $z \sim 2.5$ ultraluminous infrared starburst galaxies}

\author{Jeff~Wagg$^{1}$ and Emmanuel Momjian}

\affil{National Radio Astronomy Observatory, PO Box O, Socorro,
  NM, USA 87801\\
$^1$Max-Planck/NRAO Fellow }\email{jwagg@nrao.edu}

\begin{abstract}
We present Expanded Very Large Array and Arecibo observations of two
lensed submm galaxies at $z \sim 2.5$, in order to search for redshifted 
22.235~GHz water megamaser emission. Both SMM~J14011+0252 and
  SMM~J16359+6612 have multi-wavelength characteristics consistent
  with ongoing starburst activity, as well as CO
 line emission indicating the presence of warm molecular gas.
 Our observations do not reveal any evidence for H$_2$O megamaser emission
 in either target, while the lensing allows us to obtain deep limits 
 to the H$_2$O line luminosities, $L_{H_2O} < 7470$~L$_{\odot}$ (3-$\sigma$) in
 the case of SMM~J14011+0252, and $L_{H_2O} < 1893$~L$_{\odot}$ for
 SMM~J16359+6612, assuming linewidths of 80~km~s$^{-1}$. Our search
 for, and subsequent non-detection of 
 H$_2$O megamaser emission in two strongly lensed starburst galaxies,
 rich in gas and dust, suggests that such megamaser emission is
 not likely to be common within the unlensed population of high-redshift
starburst galaxies. We use the recent detection of strong H$_2$O megamaser
emission in the lensed quasar, MG~J0414+0534 at $z = 2.64$ to make
predictions for future EVLA C-band surveys of H$_2$O megamaser emission in
submm galaxies hosting AGN. 
  \end{abstract}


\section{Introduction}
\label{sec:intro}

 An important goal for our understanding of galaxy formation and
 evolution is to quantify the state of the molecular gas (mass, temperature,
 and density) available to form stars in young galaxies. 
 Depite many years of study at submm-to-cm wavelengths, very few observational
 probes of the molecular gas in high-redshift starburst galaxies and
 active galactic nuclei (AGN) have emerged. Most efforts
 have focused on observations of redshifted CO line emission, demonstrated 
 to be an effective tracer of the total molecular gas reservoir
 available to form stars in nearby 
 ultraluminous infrared galaxies (e.g. Downes \& Solomon 1998). 
 Searches at high-redshift have 
 successfully detected CO line emission in more than 60 galaxies 
(e.g. Solomon \& Vanden~Bout~2005), where most have been selected for
 their luminosity at far-infrared wavelengths,  
 indicating the presence of heavy elements. High dipole moment molecules
 such as HCN and HCO$^+$ have been detected in the most luminous
 objects within this CO-selected sample 
 (Solomon et al.\ 2003; Vanden~Bout et al.\ 2004; Wagg et al.\ 2005; 
 Riechers et al.\ 2006). These high-density gas tracers have
 proven very difficult to detect in high-redshift objects, as 
 they are typically an order of magnitude fainter than the
 corresponding CO line emission.

 Megamaser emission from the OH and H$_2$O molecules may serve as a
 powerful alternative means of studying the dense gas in high-redshift objects
 (see Lo 2005 for a review). 
 Generally associated with star-formation or AGN activity, 
  strong OH or H$_2$O megamaser emission could be detected at
 frequencies, $\nu \la $~10~GHz
 using current facilities. The luminosity in the OH maser line has been 
 demonstrated to correlate strongly with FIR luminosity (Darling \&
 Giovanelli 2002), while a 
 weak correlation may also exist between the FIR luminosity and that in the 22~GHz water 
 megamaser line (Castangia et al.\ 2008).  
  Surveys for both molecular species in strongly lensed quasars and
 submm galaxies (hereafter SMGs) have been conducted (Wilner et al.\
 1999; Ivison 2006; Castangia et
 al.\ 2008; Edmonds et al.\ 2009). The first detection of H$_2$O
  megamaser emission at a cosmologically significant distance was
  in a type-II QSO at $z =0.66$ by Barvainis \& Antonucci
  (2004). More recently, Impellizzeri et al.\ (2008) have detected
  H$_2$O megamaser emission in the lensed quasar, MG~J0414+0534 at
  $z=2.639$. The discovery of such megamaser emission in a quasar existing  
when the Universe was about 20\% of its present age, 
 opens up the possibility that megamaser (or kilomaser) emission may
 also be detectable in 
  equally luminous, star-forming galaxies at these early times. 
  H$_2$O masers are observed towards star-forming regions in the Milky
 Way, and indeed have been detected in a few nearby starburst galaxies 
(e.g. Haschick \& Baan 1985; Henkel, Wouterloot \& Bally 1986; 
 Hagiwara, Diamond \& Miyoshi 2002, 2003; Nakai, Sato \& Yamauchi 2002).
 Extragalactic H$_2$O megamasers with isotropic luminosities, $L_{H_2O} >
 $10~L$_{\odot}$, are confined to the nuclei of galaxies, where the
 physical conditions needed to reach population inversion, namely
 temperatures above 300~K and densities greater than 10$^7$~cm$^{-3}$,
 are more likely to be present.

Among the most luminous class of high-redshift starburst galaxies are
the submm galaxies (SMGs), discovered in blank-field
submm/mm-wavelength bolometer surveys with single-dish telescopes
 (Smail, Ivison \& Blain 1997; Hughes et al.\ 1998; Barger et al.\
1998; Bertoldi et al.\ 2000). 
These typically have 850~$\mu$m flux densities, $S_{850\mu m} \ga
5$~mJy, implying FIR luminosities, $\ga$10$^{13}$~$L_{\odot}$ (assuming a
greybody distribution with dust temperature, $T_d = 40$~K and $\beta =
1.5$ describes the FIR-to-mm emission). Optical spectroscopic 
surveys have published redshifts for less than 100 SMGs
 (e.g. Chapman et al.\ 2003, 2005). These surveys have been hindered by the 
 coarse angular resolution
of submm/mm telescopes, so that deep radio interferometry has been needed to
identify the correct multi-wavelength counterparts, and also by the faintness
of the optical/infrared counterparts (e.g. Wang et al.\ 2007). An alternative
method has been to use broad bandwidth
 mm-to-cm wavelength receivers in searching for redshifted CO line
 emission (Wagg et al.\ 2007; Daddi et al.\ 2009a, b). These searches 
 have so far been moderately successful in obtaining redshifts for $z > 3.5$ SMGs. 
 Lower frequency searches for redshifted megamaser and gigamaser emission may
 prove to be a viable alternative to searches for molecular CO line
 emission in SMGs (Townsend et al.\ 2001; Ivison et al.\ 2006; Edmonds
 et al.\ 2009), with
 telescopes like the Expanded Very Large Array (EVLA)\footnote{The National Radio Astronomy 
Observatory is a facility of the National Science Foundation operated under
cooperative agreement by Associated Universities, Inc.}.
 Such observations could also provide an effective means of probing
 the dense gas in these objects, since previous searches for HCN
 emission in SMGs have 
 not been successful (Carilli et al.\ 2005; Gao et al.\ 2007).
In this study, we present EVLA and Arecibo observations of H$_2$O
 megamaser emission in two gravitationally lensed SMGs. 
 Throughout this work, we assume H$_{0}$=71\,km\,s$^{-1}$, $\Omega$$_{M}$=0.27, and
$\Omega$$_{\Lambda}$=0.73 (Spergel et al.\ 2007).

\section{Targets}
\label{sec:targets}

The two SMGs in our sample were discovered in submm surveys of 
low redshift galaxy clusters, where gravitational lensing has been used to
constrain the faint end of the submm/mm source counts ($S_{850\mu m} < $2~mJy; 
e.g. Smail, Ivison \& Blain 1997; Knudsen et al.\ 2008). Both sources have
also been previously detected in CO line emission, so that the
emission redshift of the molecular gas region is known with 
sufficient accuracy to conduct further searches
 for redshifted H$_2$O megamaser 
emission within the narrow spectral bandwidth afforded by the current VLA
correlator.

Previous studies have observed H$_2$O megamaser emission in
high-redshift AGN, whereas our sample is comprised of purely starburst
galaxies. 
\smmjof\ at $z = 2.565$ (Barger et al.\ 1999) was discovered in the 850~$\mu$m SCUBA survey 
of the A1835 cluster field (Ivison et al.\ 2000),  
 and subsequently detected in
\cothree line emission by Frayer et al.\ (1999). The lensing
amplification factor of this source has been the subject of some
debate, and may be as high as $\sim$25 (Downes \& Solomon 2003;
Motohara 2005; Smail et al.\ 2005; Smith et al.\ 2005), however in this
analysis we adopt a lower factor of 3.5 (Smail et al.\ 2005) so as to
be conservative in our derived luminosity limits in the case of a 
 non-detection. 
 The optical spectral characteristics of this SMG are consistent with
 that of a starburst galaxy (Frayer et al.\ 1999), while the mid-IR
 spectrum is also consistent with no AGN being present 
in this SMG (Rigby et al.\ 2008). 
 Our second target, \smmjos\ is a $z=2.517$ SMG lensed into three
 components by the A2218 cluster (Kneib et al.\ 2004), with the
 brightest component lensed by a factor of $\sim$22. All three
 components have been detected in CO line emission (Sheth et al.\
 2004; Kneib et al.\
 2005), revealing a mass in molecular gas of only
 $\sim$3$\times$10$^9$~M$_{\odot}$. 
 Non-detections of  \smmjos\ in both the 0.5--2 and 2--8~keV \textit{Chandra} 
X-ray bands imply that no obscured AGN is present, 
 a conclusion supported by the mid-IR spectrum of SMM~J16359+6612, which is similar 
 to a starburst galaxy template derived from $z \sim 0$ galaxies
 (Rigby et al.\ 2008).

\section{Observations and data reduction}
\label{sec:obs}

The rest frequency of the H$_2$O maser line is 22.23508~GHz, so that
at the redshifts of our two targets ($z \sim 2.5$; Table~1), the
extended frequency coverage provided by the new EVLA C-band receivers
  (4--8~GHz) is required to conduct our search.  
Observations of \smmjos\ were obtained in two five-hour 
 tracks on September 28 and 29, 2008 (project AW750).  
 A total of 16 EVLA antennas were available during a move from the compact D to
 the most extended A configuration.
  Another five-hour track in the B
 configuration was observed toward \smmjof\ on February 17, 2009
 (project AW761), when
 18 EVLA antennas were available. In both cases, the
 central tuning frequency was dictated by the 
redshift of the \hij CO lines detected previously in these two
 sources (Frayer et al. 1999; Sheth et al.\ 2004). A bandwidth of
 6.25~MHz was used to cover the velocity range of the CO
lines, and a spectral resolution of 97.656~kHz was adopted.

Flux density and bandpass calibration were performed using
observations of 3C286, while
 we observed 1642+689 as a phase calibrator for SMM~J16359+6612, and
  1354-021 for SMM~J14011+0252. 
The data were reduced using the NRAO's Astronomical Image Processing System (AIPS). 
 Table~1 summarizes our EVLA observation parameters.

We have also observed  J14011+0252 with the 305~m Arecibo radio 
telescope\footnote{The Arecibo Observatory is part of the National
  Astronomy and Ionosphere Center, which is operated by Cornell
  University under a cooperative agreement with the National Science
  Foundation.}, in order to search for H$_2$O 
megamaser emission at a frequency of 6.237~GHz. These data were obtained during
3.5 hours of observing on March 12 and 13, 2009, with the C-high
receivers and using the standard
`ON/OFF' observing mode. Both orthogonal polarizations were recorded.
  Five minutes were spent in
each of the `ON' and `OFF' positions during a single scan. The correlator was
configured to provide 64 channels of spectral resolution, and the final rms per
channel is 152~$\mu$Jy/beam after hanning smoothing. These
observations are $\sim$20\% more sensitive than our EVLA observations
of the same source. The data were analyzed using standard Arecibo
Observatory \textit{idl} routines.

\section{Results}
\label{sec:results}

\begin{center}
\begin{figure}
\includegraphics[scale=0.7]{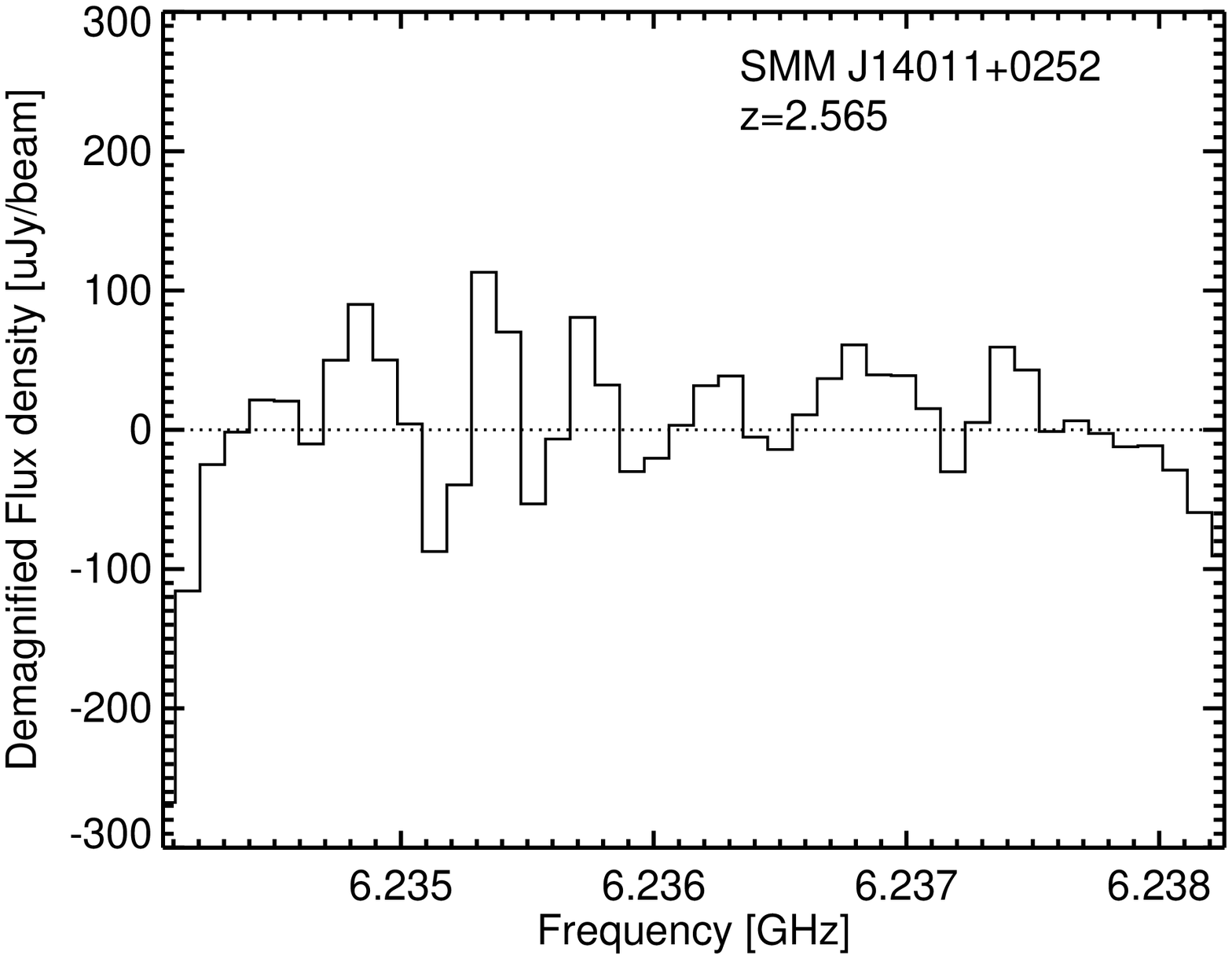}

\includegraphics[scale=0.7]{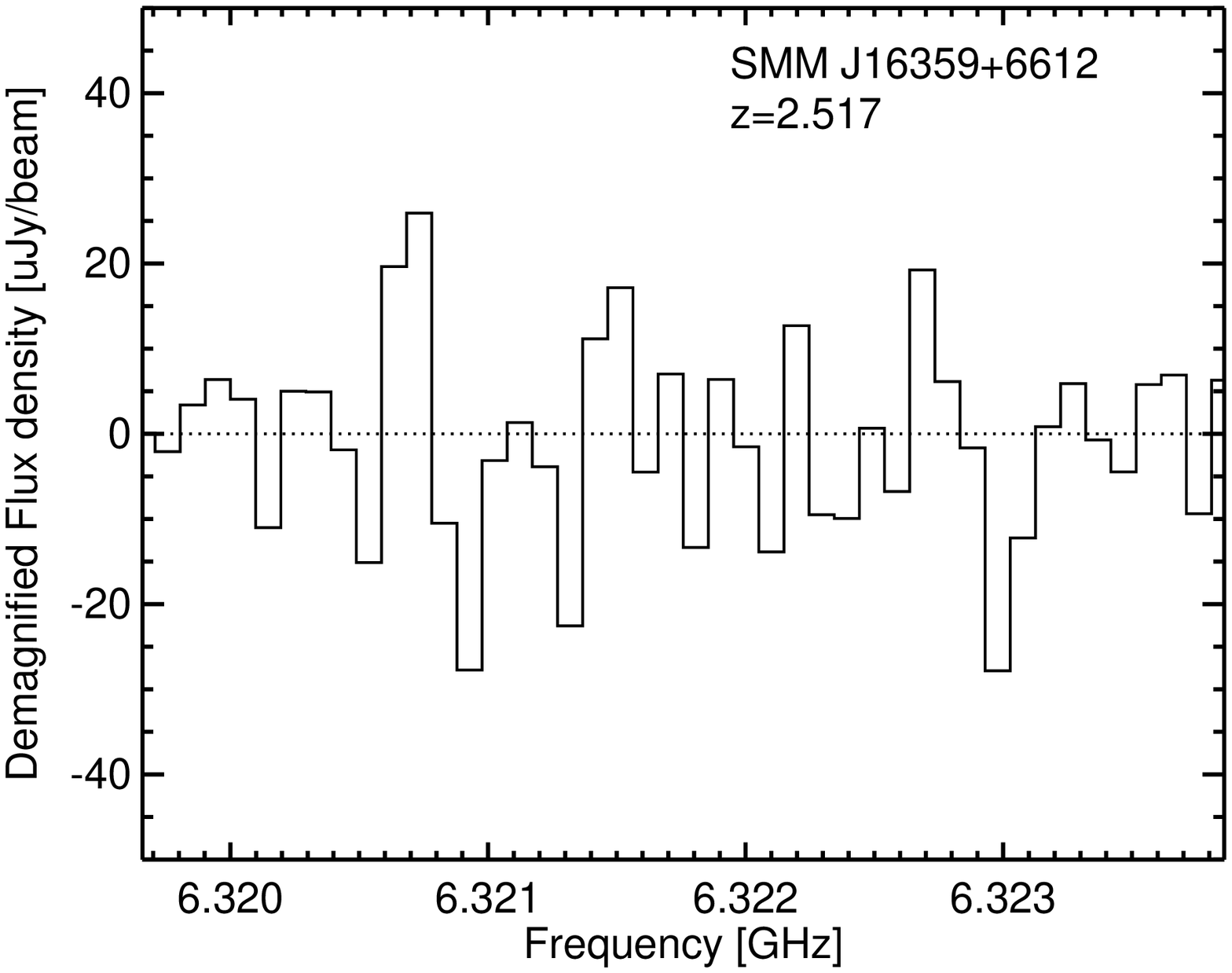}

\caption{\textit{top:} C-band high Arecibo spectrum of SMM~J14011+0252 covering
  6.25~MHz with a spectral resolution of 97.7~kHz and corrected for
  amplification by gravitational lensing.  
\textit{bottom:} Spectrum of the C-band emission from the 3 lensed
components of  
SMM~J16359+6612, averaged after correcting each for the appropriate
magnification factor. The original bandwith used is 6.25~MHz with a
spectral resolution of 97.7~kHz.}
\label{fig:fig1}
\end{figure}
\end{center}

 Our C-band spectra of \smmjof\ and \smmjos\
 do not reveal any evidence for H$_2$O line emission. Given the strong
 lensing amplification of these objects by foreground galaxy clusters, 
 our spectra can be used to derive very deep limits to the line luminosities of any 
 H$_2$O megamasers which may be present. Following Edmonds et al.\
 (2009), we first correct the spectrum extracted at each of the three
 components of \smmjos\
 for its corresponding lensing factor (Kneib et al.\ 2005),
 before adding these together with the appropriate weighting to 
 create the average spectrum shown in Figure~\ref{fig:fig1}. The rms
 of this average spectrum is 11.6$\mu$Jy per 97.7~kHz channel, which is a
 factor of $3\times$ lower than that obtained by Edmonds et
 al.\ (2009). We decide to use the more 
  sensitive Arecibo spectrum in calculating our H$_2$O line
 luminosity limits, which is plotted in  Figure~\ref{fig:fig1}.
 The lensing-corrected rms of this 
 Arecibo spectrum is 43.4$\mu$Jy per 97.7~kHz.

Our line luminosity limits are calculated for a range in  H$_2$O
megamaser linewidths, 20, 40, 60, and 80~km~s$^{-1}$, consistent with that of 
nearby FIR luminous galaxies (Henkel et al.\ 2005). We then calculate
lensing-corrected 3-$\sigma$ limits to the isotropic line
luminosity in SMM~J16359+6612, $L_{H_2O} <$~946, 1339, 1639, and
1893~$L_{\odot}$. Similarly, in \smmjof\ the isotropic line luminosity
limits are $L_{H_2O} <$~3735, 5282, 6469, and 7470~$L_{\odot}$ for our
assumed linewidths of 20, 40, 60 and 80~km~s$^{-1}$.

\section{Discussion}
\label{sec:discuss}

A previous search for H$_2$O megamaser emission in \smmjos\ has been 
conducted by Edmonds et al.\ (2009). While the sensitivity achieved 
here is nearly a factor of three higher, we do not find
any evidence for line emission. Assuming a dust temperature, $T_d =
40$~K, the corrected 850~$\mu$m flux density of \smmjos\ is consistent
with a FIR luminosity, $L_{FIR} \sim 1.5 \times 10^{12}$~L$_{\odot}$,
which can be used to predict the expected H$_2$O megamaser line
luminosity according to the relationship observed in star-forming
regions (Genzel \& Downes 1979; Jaffe et al. 1981). The predicted line
luminosity is, $L_{H_2O} \sim 1500$~L$_{\odot}$, consistent with our
derived limits. In the case of SMM~J14011+0252, the corrected 850~$\mu$m
flux density would imply, $L_{FIR} \sim 6.7 \times
10^{12}$~L$_{\odot}$, for the same assumptions on the dust temperature. 
We should then expect an H$_2$O megamaser luminosity, 
$L_{H_2O} \sim 6700$~L$_{\odot}$, which is also consistent with our
observed limits. If we assume the weak correlation between the 
H$_2$O megamaser luminosity and FIR luminosity measured for FIR
bright galaxies by Castangia et al.\ (2008), we would expect 
megamaser line luminosities of, 3800 and 16800~L$_{\odot}$ for \smmjos and
SMM~J14011+0252, respectively.

The sensitivity of our constraints on H$_2$O megamaser emission in these two
star-forming ultraluminous infrared galaxies (ULIRGs) would suggest
that they do not contain strong megamasers, such as that observed in 
MG~J0414+0534 at $z = 2.639$. Although our sample size is
small, these non-detections in strongly lensed, metal and gas-rich
starburst galaxies would imply that strong H$_2$O megamaser emission may not
be common among such objects. As pointed out by Edmonds et al.\ (2009), it
is likely that the physical conditions of the interstellar gas in
these predominantly star-forming SMGs is not sufficiently warm and
dense to excite megamaser line
emission, which is more readily excited in the X-ray irradiated
 environments close to a central AGN (Neufeld et al.\ 1994). Another
 possibility is that the density and temperature of the gas are large, 
 but the gas is not sufficiently coherent in velocity 
 with the maser source to produce strong amplification.

Weiss et al.\ (2005) observe multiple \hij transitions of CO line emission
in SMM~J16359+6612, which reveal a double-peaked CO line profile,
consistent with either a circumnuclear toroid of rotating molecular gas, or 
molecular gas associated with separate components of an ongoing
merger. Their excitation analysis finds the gas to be cool, with
temperatures most likely below $\sim$80~K, while the large 
star-formation rate ($\sim$500~M$_{\odot}$~yr$^{-1}$) inferred by 
Kneib et al.\ (2004) should 
result in a high gas excitation if the CO line emission were due to a
single, massive component. This low excitation gas spread over multiple
components is also consistent with our non-detection of an H$_2$O
megamaser in SMM~J16359+6612.

Although our luminosity limits are consistent with what
one would expect from a simple extrapolation of the
$L_{FIR}-to-L_{H_2O}$ relation in 
 Galactic star-forming regions, these limits do not rule
out the presence of  H$_2$O ``kilomasers'' ($L_{\odot} <
10$~L$_{\odot}$), which are typically found 
 both within, and exterior to the nuclei of nearby star-forming galaxies 
 (e.g. Hagiwara et al. 2001, 2003; Henkel et al. 2004). Detecting such
 emission in a member of the, $S_{850\mu m} > $5~mJy, SMG population 
 would require hundreds of hours of observing time using the full
 EVLA, or Arecibo. Nearby ulitraluminous infrared galaxies, which are
 the nearest analogues to the high-redshift SMGs, generally do not
 exhibit strong water megamaser emission. The exceptions to this
 are NGC6240 and UGC5101 (Hagiwara et al. 2002; Zhang et al. 2006),
 which are both believed to contain AGN. Such an association of strong
 H$_2$O megamaser emission with nearby AGN, rather than starburst
 ULIRGs, is consistent with our non-detections in \smmjos and
 SMM~J14011+0252. This would also support the original
 claim by Edmonds et al.\ (2009) that H$_2$O megamasers will not be an
 effective means of measuring redshifts for starburst SMGs in future EVLA
 surveys.

\subsection{Predictions for future EVLA surveys}

Although we do not detect megamaser emission in the two starburst
galaxies observed, the detection of an H$_2$O megamaser in
MG~J0414+0534 suggests that such emission may also be present in SMGs
hosting an AGN. The results of deep X-ray imaging of SMGs shows that 
 28--50\% may
harbour an AGN (Alexander et al.\ 2005), while the median mid-infrared
spectrum of a sample of 24 SMGs (Men\'endez-Delmestre et al.\ 2009), 
 suggests that less than one third of
 the FIR luminosity is likely powered by such AGN
 (Pope et al.\ 2008). 
 We can use these results to make
simple predictions for future C-band surveys with the EVLA.

With the already existing, expanded frequency coverage of the
EVLA C-band receivers, it is possible to search for water megamaser emission in $z >
1.78$ SMGs.  The forthcoming availability of the WIDAR correlator
will also make it possible to obtain spectroscopy over the entire 4--8~GHz
window, so that searches for redshifted megamaser emission could be
conducted over the redshift range, $1.78 < z < 4.56$, within a field
of view of 44~sq. arcmins in size. The best current constraints on the
850~$\mu$m number counts predict about 0.1 SMGs with $S_{850\mu m} >
$6~mJy, per sq. arcmin (Coppin et al.\ 2006), so that we would expect
4.4 bright SMGs in a single C-band pointing. Assuming $T_d = 40$~K and $\beta
= 1.5$ for this sub-sample of the population, then their expected FIR
luminosities would be, $L_{FIR} > 1.1\times 10^{13}$~L$_{\odot}$. 
 Roughly 73\% of the bright, radio-detected SMG population (Chapman et
 al.\ 2005), would have a redshift that places the 22.23508~GHz water
 maser line in the frequency interval covered by the EVLA C-band. 
 While it may be that the 15-20\% of bright SMGs with no 1.4~GHz radio
 counterpart are at redshifts, $z >> 3.5$
(e.g. Ivison et al.\ 2005; Wagg et al.\ 2009), 
 we will assume here that all of these are at
 redshifts, $z < 4.56$ (although see Coppin et al.\ 2009). 
 If we make the assumption that 30\% of the SMGs in this
 redshft range have AGN, and that 30\% of their FIR luminosity is
 powered by this AGN, then we would expect one SMG in a single C-band
 pointing to have AGN with FIR luminosities, $L_{FIR} > 3.8\times
 10^{12}$~L$_{\odot}$. If we assume the ratio between $L_{FIR}$
 and L$_{H_2O}$ measured in MG~J0414+0534 (Impellizzeri et al.\ 2008), and
 also that these SMGs do indeed host a water megamaser
 where the accretion disk is being viewed edge-on, then
 they could exhibit isotropic H$_2$O line luminosities,
 $L_{H_2O} > 40000$~L$_{\odot}$. For the faintest, unlensed sources 
 at $z = 3.5$, a maximum of 30 EVLA hours would be needed to
 detect the megamaser emission. However, we note that no unlensed
 megamaser has been detected with such an extreme luminosity, and it
 is possible that a cutoff in the luminosity function of H$_2$O
 megamasers exists at high luminosities. 
 Based on their detection in MG~J0414+0534, 
 Impellizzeri et al.\ (2008) argue that
 the probability of finding an H$_2$O megamaser in a single, pointed
 observation of a high-redshift object is 10$^{-6}$ for a non-evolving
 line luminosity function, and 0.05 if this function evolves strongly
 with redshift, like (1 + $z$)$^4$. Therefore, ``blind'' EVLA surveys
 for redshifted megamaser emission will serve to constrain any
 evolution in the luminosity function.

\section{Summary}
\label{sec:sum}

We have used the EVLA and Arecibo to search for H$_2$O megamaser
emission in two strongly lensed, starburst SMGs at $z \sim 2.5$. The
gravitational lensing of these two objects allows us to rule out the
presence of strong megamaser emission, such as that observed in
 MG~J0414+0534 at $z = 2.639$. We are not able to rule out the
 presence of kilomaser emission commonly
associated with star-forming galaxies. These are the deepest
 limits to date on the luminosities of H$_2$O megamasers in
 high-redshift galaxies.

It is possible that future C-band
surveys with the EVLA will detect one H$_2$O megamaser emitting SMG
within a single, EVLA pointing. Currently planned
EVLA observations of SMGs with observed AGN characteristics, and
existing CO emission line data, will allow us to better assess the
success rate of conducting such ``blind'' megamaser surveys in the future. 
At the time of completion, the EVLA will have been
outfitted with new X-band (8-12~GHz) and Ku-band receivers
(12-18~GHz), so that it will also be possible to conduct searches for
H$_2$O megamasers at lower redshifts.

\section{Acknowledgments}

We thank the NRAO staff involved in
the EVLA project for making these observations possible. 
JW is grateful for support from the Max-Planck Society and the
Alexander von Humboldt Foundation. We would also like to 
 thank David Wilner, Chris Carilli, Karl
Menten, Liz Humphries, David Hughes and Robert Edmonds
for helpful discussions.

\begin{deluxetable}{lcc}
\tablecaption{H$_2$O observations of SMM J16359+6612 and SMM~J14011+0252.
\label{table1}}
\tablewidth{0pt}
\tablehead{\colhead{} & \colhead{SMMJ16359} & \colhead{SMMJ14011}}
\startdata
$z_{CO}$: &      2.517      &   2.565  \\
Frequency:& 6.3221~GHz   &  6.2365~GHz \\
Pointing center (J2000) RA: & 16$^h$35$^m$44$^s$.15 & 14$^h$01$^m$04$^s$.92 \\
                       Dec: & +66$^o$12$^m$24$^s$.0  &  +02$^o$52$^m$25$^s$.6 \\
Synthesized beam: & $1\farcs70 \times 1\farcs58$, P.A. 36.4$^{\circ}$ & $1\farcs80 \times 1\farcs28$, P.A. $5.8^{\circ}$ \\
Channel spacing: & 97.7~kHz & 97.7~kHz \\
line rms (97.7~kHz channel):   & 221~$\mu$Jy~beam$^{-1}$ & 275~$\mu$Jy~beam$^{-1}$  \\
\enddata            
\end{deluxetable}


\begin{thebibliography}{}

\bibitem{alexander05} Alexander, D.M. et al.\ 2005, Nature, 434, 738

\bibitem{Barger98} Barger A.J., Cowie L.L., Sanders D.B., Fulton E., Tanigushi Y., Sato Y., Kawara K., Okuda H., 1998, Nat, 394, 248

\bibitem[Barger et al.(1999)]{1999AJ....117.2656B} Barger, A.~J., Cowie, 
L.~L., Smail, I., Ivison, R.~J., Blain, A.~W., 
\& Kneib, J.-P.\ 1999, \aj, 117, 2656 

\bibitem{2005ApJ...628L..89B} Barvainis R., Antonucci R., 2005, ApJ, 628, L89 

\bibitem{bertoldi02} Bertoldi, F., et al. 2000, A\&A, 360, 92


\bibitem[Carilli et al.(2005)]{2005ApJ...618..586C} Carilli, C.~L., et al.\ 
2005, \apj, 618, 586 

\bibitem[Castangia et al.\ 2008]{castangia08} Castangia, P. et al.\ 2008, A\&A, 479, 111C

\bibitem{chapman05} Chapman S.C., et al.\ , 2005, ApJ, 622, 772


\bibitem[Coppin et al.(2006)]{2006MNRAS.372.1621C} Coppin, K., et al.\ 
2006, \mnras, 372, 1621 

\bibitem[Coppin et al.(2009)]{2009arXiv0902.4464C} Coppin, K., et al.\ 
2009, arXiv:0902.4464 



\bibitem[Daddi et al.(2009)]{2009ApJ...695L.176D} Daddi, E., Dannerbauer, 
H., Krips, M., Walter, F., Dickinson, M., Elbaz, D., 
\& Morrison, G.~E.\ 2009a, \apjl, 695, L176 

\bibitem[Daddi et al.(2009)]{2009ApJ...694.1517D} Daddi, E., et al.\ 2009b, 
\apj, 694, 1517 



\bibitem{darling02} Darling, J., \& Giovanelli, R. 2002, 124, 100

\bibitem[Downes \& Solomon(1998)]{1998ApJ...507..615D} Downes, D., \& Solomon, P.~M.\ 1998, \apj, 507, 615 

\bibitem{downes03} Downes, D., \& Solomon, P. M. 2003, ApJ, 582, 37 

\bibitem[Edmonds et al.(2009)]{2009AJ....137.3293E} Edmonds, R., Wagg, J., 
Momjian, E., Carilli, C.~L., Wilner, D.~J., Humphreys, E.~M.~L., Menten, 
K.~M., \& Hughes, D.~H.\ 2009, \aj, 137, 3293 

\bibitem[Frayer et al.(1999)]{1999ApJ...514L..13F} Frayer, D.~T., et al.\ 1999, \apjl, 514, L13 

\bibitem{Gao07} Gao, Y., et al.\ 2007, \apj, 660, L93

\bibitem{1979A&A....72..234G} Genzel, R., \& Downes, D.\ 1979, \aap, 72, 234 


\bibitem{hagiwara01} Hagiwara, Y., Henkel, C., Menten, K. M., \& Nakai, N. 2001, ApJ, 560, L37

\bibitem{hagiwara02} Hagiwara Y., Diamond P.J., Miyoshi M., 2002, A\&A, 383, 65

\bibitem{hagiwara03} Hagiwara Y., Diamond P.J., Miyoshi M., 2003, A\&A, 400, 457 

\bibitem{Haschick85} Haschick, A.D. \& Baan W.A., 1985. \nat 314, 144

\bibitem{henkel86} Henkel C., Wouterloot J.G.A., Bally J., 1986, A\&A 155, 193

\bibitem{henkel04} Henkel, C., Tarchi, A., Menten, K. M., \& Peck, A. B. 2004, A\&A, 414, 117

\bibitem{henkel05} Henkel, C., et al.\ 2005, A\&A, 436, 75

\bibitem{Hughes98} Hughes D.H., et al., 1998, Nat, 394, 241

\bibitem[Impellizzeri et al.(2008)]{2008Natur.456..927I} Impellizzeri, C.~M.~V., McKean, J.~P., 
Castangia, P., Roy, A.~L., Henkel, C., Brunthaler, A., \& Wucknitz, O.\ 2008, \nat, 456, 927 

\bibitem[Ivison et al.(2000)]{2000MNRAS.315..209I} Ivison, R.~J., Smail, 
I., Barger, A.~J., Kneib, J.-P., Blain, A.~W., Owen, F.~N., Kerr, T.~H., 
\& Cowie, L.~L.\ 2000, MNRAS, 315, 209 


\bibitem{ivison05} Ivison, R.J. et al.\ 2005, MNRAS, 364, 1025

\bibitem{ivison06} Ivison R.~J., 2006, MNRAS, 370, 495


\bibitem{jaffe81} Jaffe, D. T., Guesten, R., \& Downes, D. 1981, ApJ, 250, 621

\bibitem[Kneib et al.\ 2004]{kneib04} Kneib, J.-P., et al.\ 2004, MNRAS, 349, 1211

\bibitem[Knudsen et al.(2008)]{2008MNRAS.384.1611K} Knudsen, K.~K., van der 
Werf, P.~P., \& Kneib, J.-P.\ 2008, \mnras, 384, 1611 

\bibitem{2005ARA&A..43..625L} Lo, K.~Y.\ 2005, ARA\&A, 43, 625 

\bibitem{2009arXiv0903.4017M} Men{\'e}ndez-Delmestre, K., et al.\ 2009, arXiv:0903.4017 

\bibitem{motohara05} Motohara, K., et al. 2005, AJ, 129, 53

\bibitem{nakai02} Nakai N., Sato N., Yamauchi A., 2002. ApJ, 54, L27

\bibitem[Pope et al.(2008)]{2008ApJ...675.1171P} Pope, A., et al.\ 2008, 
\apj, 675, 1171 


\bibitem[Riechers et al.(2006)]{2006ApJ...645L..13R} Riechers, D.~A., 
Walter, F., Carilli, C.~L., Weiss, A., Bertoldi, F., Menten, K.~M., 
Knudsen, K.~K., \& Cox, P.\ 2006, \apjl, 645, L13 

\bibitem[Sheth et al.(2004)]{2004ApJ...614L...5S} Sheth, K., Blain, A.~W., 
Kneib, J.-P., Frayer, D.~T., van der Werf, P.~P., 
\& Knudsen, K.~K.\ 2004, \apjl, 614, L5 

\bibitem{Smail97} Smail, I., Ivison R.J., Blain A.W., 1997, ApJ, 490, L5

\bibitem{smail05} Smail, I., Smith, G. P., \& Ivison, R. J. 2005, ApJ, 631, 121

\bibitem{smith05} Smith, G. P., et al.\ 2005, MNRAS, 359, 417

\bibitem[Solomon et al.(2003)]{2003Natur.426..636S} Solomon, P., Vanden 
Bout, P., Carilli, C., \& Guelin, M.\ 2003, \nat, 426, 636

\bibitem[Solomon \& Vanden Bout 2005]{solomon05} Solomon, P.M., \& Vanden Bout, P. A. 2005, ARA\&A, 43, 677

\bibitem{townsend01} Townsend, R.H.D., et al.\  2001, MNRAS, 328, L17

\bibitem[Vanden Bout et al.(2004)]{2004ApJ...614L..97V} Vanden Bout, P.~A., 
Solomon, P.~M., \& Maddalena, R.~J.\ 2004, \apjl, 614, L97 

\bibitem[Wagg et al.(2005)]{2005ApJ...634L..13W} Wagg, J., Wilner, D.~J., 
Neri, R., Downes, D., \& Wiklind, T.\ 2005, \apjl, 634, L13 

\bibitem{wagg07} Wagg, J., et al.\ 2007, MNRAS, 375, 745

\bibitem[Wagg et al.(2009)]{2009arXiv0905.0691W} Wagg, J., Owen, F., 
Bertoldi, F., Sawitzki, M., Carilli, C.~L., Menten, K.~M., 
\& Voss, H.\ 2009, AJ, 137, 3293

\bibitem{wang07} Wang, W.-H., Cowie, L.~L., van Saders, J., Barger, A.~J., \& Williams, J.~P.\ 2007, \apjl, 670, L89 

\bibitem[Wei{\ss} et al.(2005)]{2005A&A...440L..45W} Wei{\ss}, A., Downes, D., Walter, F., \& Henkel, C.\ 2005, \aap, 440, L45 

\bibitem[Wilner et al.\ 1999]{wilner99} Wilner D. J. et al.\ 1999, \apj, 117, 1139

\bibitem[Zhang et al.(2006)]{2006A&A...450..933Z} Zhang, J.~S., Henkel, C., Kadler, M., Greenhill, L.~J., Nagar, N., Wilson, A.~S., \& Braatz, J.~A.\ 2006, \aap, 450, 933 




\end{thebibliography}
\end{document}